\documentclass[10pt,sigconf,letterpaper]{acmart}
\renewcommand\footnotetextcopyrightpermission[1]{} 
\usepackage[T1]{fontenc}
\usepackage{graphicx}
\usepackage{hyperref}
\usepackage{xcolor}
\usepackage[font=small,labelfont=bf, textfont=it]{caption}
\usepackage[hang,scriptsize,tight,nooneline]{subfigure}

\newcommand{\parab}[1]{\vspace{0.03in}\noindent{\bf #1}}
\newcommand{\starlinksubreddit}{\texttt{r/Starlink}}

\newcommand{\aryan}{\textcolor{black}}

\begin{document}
%
\title{\fontsize{18pt}{20pt}\selectfont On viewing SpaceX Starlink through the Social Media Lens}
%
%
%
%
%
%
\author{Aryan Taneja, Debopam Bhattacherjee, Saikat Guha, Venkata N. Padmanabhan}
\affiliation{%
\vskip 0.5em
  \institution{Microsoft Research - India}
\vskip 1.5em
}

\begin{abstract}
\textit{Multiple low-Earth orbit satellite constellations, aimed at beaming broadband connectivity from space, are currently under active deployment. While such space-based Internet is set to augment, globally, today's terrestrial connectivity, and has managed to generate significant hype, it has been largely difficult for the community to measure, quantify, or understand the nuances of these offerings in the absence of a global measurement infrastructure -- the research community has mostly resorted to simulators, emulators, and limited measurements till now. In this paper, we identify an opportunity to use the social media `lens' to complement such measurements and mine user-centric insights on the evolving ecosystem at scale.}

\textit{To illustrate the broader opportunity here, we focus on SpaceX Starlink -- a mega-constellation that is set to eventually include tens of thousands of Low-Earth Orbit (LEO) satellites providing Internet connectivity from space. Tapping social media, we present our analyses on learning about Starlink's network events even \textit{weeks before} any public announcements on these, how some large-scale events have significantly impacted user sentiment, and how users' \aryan{perception} 
of Starlink performance has evolved over time. We discuss how the methodology presented here could enable a better, holistic understanding of these networks.}

\end{abstract}
\maketitle 

\section{Introduction}
\label{sec:intro}

Low-Earth Orbit (LEO) satellite networks like SpaceX Starlink~\cite{starlink_40K}, OneWeb~\cite{oneweb}, Amazon's Kuiper~\cite{amazon_news}, Telesat's Lightspeed~\cite{telesat}, and others are set to revolutionize the landscape of global connectivity. Multiple of these LEO satellite constellations are currently under deployment and SpaceX has already started offering Starlink Internet services in $58$ countries~\cite{beta_500K} as of June, $2023$. They already have more than $4$,$000$~\cite{starlink_count_new} satellites deployed and their long-term plan~\cite{starlink_40K} is to deploy $40K+$ LEO satellites. Recent advances in launching~\cite{reusable_boosters,10x_launch_cost_reduction} and satellite design~\cite{satellite_size} have enabled such large-scale deployments consisting of thousands of satellites. While such deployments have managed to generate significant hype~\cite{leo_popular_1,leo_popular_2,leo_popular_3,leo_popular_4,leo_popular_5}, 
there is still a serious dearth of tools that could measure these networks and quantify their performance at scale. Although the first steps have been taken to simulate~\cite{kassing2020exploring} and emulate~\cite{lai2023starrynet} LEO networks and there have been reports on limited network measurement experiments~\cite{starlink_perf_firstlook,apnic_leo_cc}, the broader community, which includes potential customers, ISPs (read competitors), researchers, and Internet enthusiasts, is still in search of a platform that gives a holistic picture of this `space'. In this work, we explore if relevant information readily available on various social media could be mined to fill this gap in understanding LEO networks.

Social platforms like Facebook, Reddit, LinkedIn, Instagram, and even newer ones like Discord have seen large increases~\cite{facebook_stat,discord_stat} in user base and online communities over the last decade. Reddit, which is a leading platform for topic-based discussion, now hosts more than $100$,$000$ active communities and generates more than $350$ million posts a year~\cite{reddit_stat}. Such tremendous growth of social media has also driven, to a large extent, a lot of active research on sentiment analysis and opinion mining. As more users share their personal experiences and express their feelings online, such language capabilities could be leveraged to distill subjective product feedback across online forums~\cite{liu2012sentiment,anto2016product}. While on one hand, the language capabilities have improved significantly~\cite{devlin2018bert,dai2019transformer,yang2019xlnet,liu2019roberta,qi2020stanza,azure_acs,google_NL_AI} in the last few years, on the other hand, these tools have also been applied to diverse sentiment analysis and opinion mining use cases -- predict election results and box-office collections of movies, understand attitude towards vaccination, detect hate speech, etc. Computer vision has seen comparable advances, and today we have document scanning and parsing (OCR) capabilities on our mobile phones~\cite{google_lens,ms_lens}. We use these new-age language and vision capabilities and other techniques to analyze users' perception of SpaceX Starlink on a popular subreddit \starlinksubreddit{}\cite{starlink_subreddit}. We believe our techniques could be generalized across LEO providers and social platforms to gather useful insights into the LEO broadband space.

\begin{figure*}[tbh]
        \begin{center} \includegraphics[width=\textwidth]{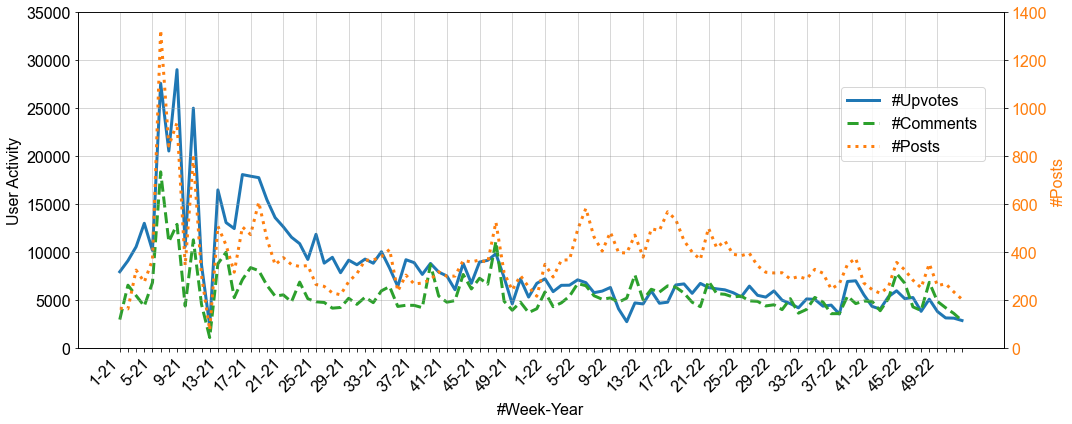}
                \vspace{-0.2in}
                \caption{Activity on \starlinksubreddit. The left $y$-axis corresponds to both the number of upvotes and comments.
                }
                \label{fig:activity_plot}
        \end{center}
\end{figure*}

Following the footsteps of past work on mining social platforms to quantify network performance and demands~\cite{qiu2010listen,hsu2011using,yang2016estimating}, and network and service failures~\cite{motoyama2010measuring,takeshita2015early}, we demonstrate here the scope of building a framework that enables the community around the `LEO' space to understand these networks better. \starlinksubreddit{} has managed to draw significant participation from enthusiasts, early adopters, and others. As, Fig.~\ref{fig:activity_plot} shows, there are $372$ posts per week on average with the peak reaching $1$,$326$ posts/week in Feb'$21$. The number of upvotes and comments, which are strong signals of user activity, are $8$,$190$ and $5$,$702$ per week on average respectively. While the initial flurry of enthusiasm (first half of $2021$) has subsided, the subreddit continues to see significant user engagement. We analyze the publicly available posts on the subreddit channel and gather aggregated insights on users' perception of Starlink service. We gather data using standard social network APIs and use standard language and computer vision tools, coupled with a set of custom heuristics, to extract structured data on network performance from publicly shared screenshots, to generate word clouds of $n$-grams, and to quantify user sentiment. Armed with this data, we explore the following $3$ representational use cases in this paper:

\vspace{-0.2in}
\begin{itemize}
    \item We analyze how user sentiment follows related events and public announcements -- this could be used by competing ISPs and the research community to evaluate the large-scale impact of changes and events -- both technical and otherwise. Also, such sentiment insights could equip potential Starlink customers to take informed decisions while subscribing to broadband services. We also show how specific events like Starlink network outages could be detected from public discussions on social media leveraging a mix of language and sentiment analysis tools.
    \item Using publicly available information and a combination of language tools, we could identify when roaming was enabled as a service on the Starlink network, weeks before any public announcement on this from Starlink. For competing ISPs, it is important to detect such signals early and act upon them. 
    \item We analyze how bandwidth on the Starlink network varies relatively over time with more launches and users, and how that drives user sentiment. Such insights could help the broad community to understand how customer expectation changes as Starlink services evolve.
\end{itemize}
\vspace{-0.05in}

We believe our analyses make the case for coming up with a framework encompassing multiple social platforms (Reddit is just one of them) and LEO broadband providers (Starlink is just one of them) that help augment the currently limited LEO measurement landscape. While here we present early-stage analyses, this framework has the potential to offer a \textbf{complementary user-centric view} of the LEO broadband space even in the presence of large-scale LEO measurement platforms in the future.

\section{Methodology}
\label{sec:methods}

In this section, we briefly discuss the APIs and language and vision tools we use to collect, extract, and analyze publicly available information on the Reddit social platform. We also describe the data we analyze.

We used Reddit~\cite{reddit_api} and pushshift.io~\cite{pushshift_api} APIs to get all the post and comment data available on \starlinksubreddit{} for the entire period between Jan'$21$ and Dec'$22$ ($24$ months). We used the Python Reddit API Wrapper (PRAW)~\cite{praw} to access the Reddit APIs. While gathering the data, we cleaned it to get rid of user IDs and content that have been explicitly requested to be removed by users. The former step makes sure that our work does not raise any ethical concerns while the latter step conforms to data privacy guidelines. We only collected data available \texttt{publicly} on the Reddit platform.

We use the following language and computer vision tools for the analyses presented in this paper:
\begin{itemize}
    \item Azure's Cognitive Services~\cite{azure_acs}: We use sentiment analysis and opinion mining APIs to quantify the sentiment of posts and comments. We also use ACS' optical character recognition (OCR) service to extract network measurement information from screenshots of speed-test reports shared by Redditors (Reddit users) on the platform.
    \item Natural Language Toolkit (NLTK)~\cite{nltk}: We use the libraries to filter out stop-words, extract $n$-grams, and create word clouds of posts and comments.
\end{itemize}

Rather than reinventing the wheel,
we rely on the already available language and vision tools. Our focus is on demonstrating how the capabilities of these tools could be leveraged to understand users' perceptions of networks better. The accuracy of our analyses depends on the accuracy of these tools.

\begin{figure}[t]
  \centering
  \subfigure[]{\label{fig:ookla1}
    \includegraphics[width=0.8\columnwidth]{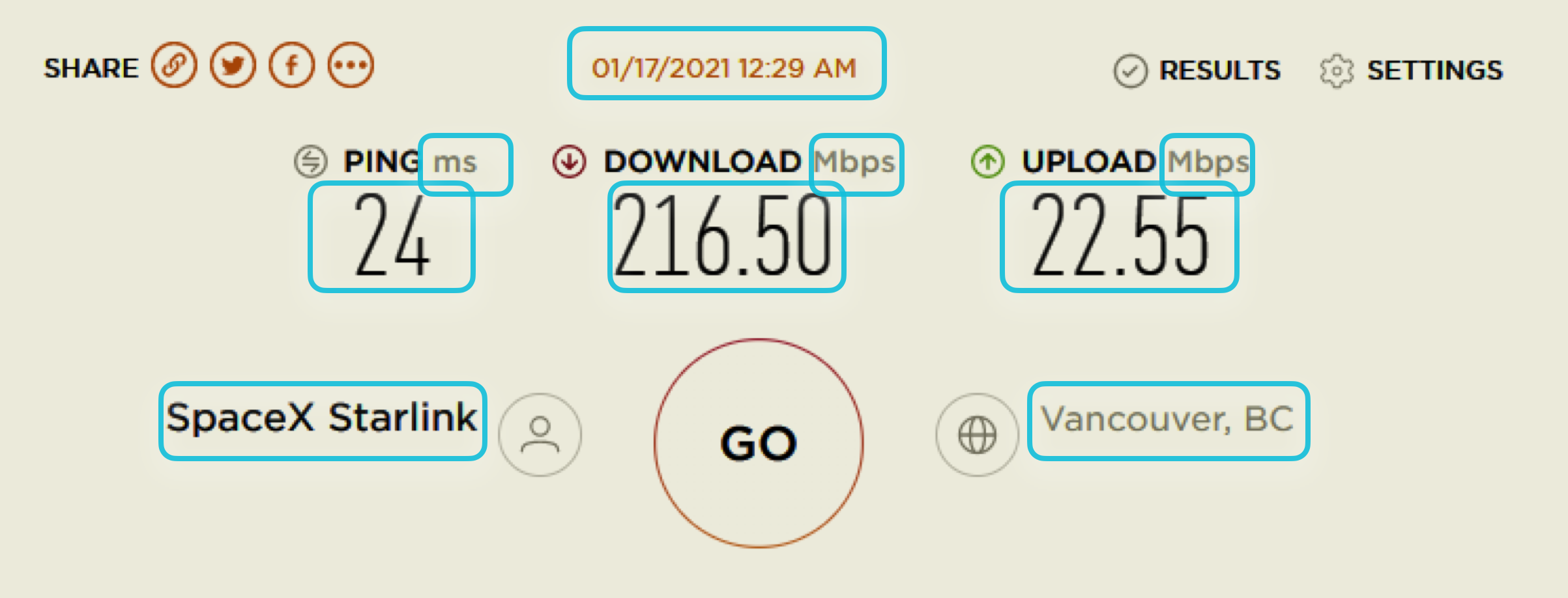}}
  \subfigure[]{\label{fig:ookla2}
    \includegraphics[width=0.6\columnwidth]{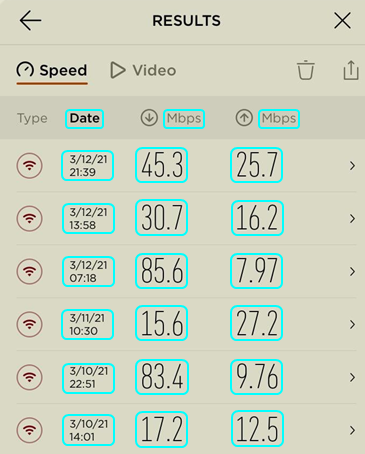}}
  \vspace{-0.15in}
  \caption{(a) Simple and (b) table-structured screenshots (standard negative of the original) of speed-test reports from Ookla, posted on \starlinksubreddit{}. Boxes with red borders mark data that could be extracted.}
  \label{fig:ookla_template}
\end{figure}

Nevertheless, we built a set of heuristics to extract network measurement metrics from the OCR output of semi-structured speed-test reports across all test providers. Fig.~\ref{fig:ookla_template} show two such templates from Ookla. Our relative pixel-distance-based heuristics could extract downlink and uplink speeds and latency (also jitter and packet loss rates, when available) along with the corresponding units, date and time information, network provider name, server location, etc. While we do not use speed-test location information in our current analyses, we could have used the server location as a proxy for client location, depending on the use case, as test providers like Ookla (present at $1$,$000+$ locations~\cite{ookla_server_sel}) usually pick a `closest'-on-the-network server for the measurements. The heuristics could parse all the different speed-test templates, including more complex table-structured ones (having multiple sub-reports together, like in Fig.~\ref{fig:ookla2}), which we could collect over the entire period. We discuss the related results in \S\ref{sec:bandwidth}.

While the readers already have a rough idea of the data by now, let us briefly list down the inputs to our analysis framework for clarity:
\begin{itemize}
    \item Reddit posts: We gather post text, embedded URLs, photos, and videos, submission timestamp, post ID, and post metadata (number of comments and upvotes).
    \item Comments: We collect comment text, embedded URLs, photos, and videos, submission timestamp, comment ID, and metadata. We regenerate the comment tree for a post leveraging parent IDs in the comment metadata.
    \item Speed-test screenshots: Many people post screenshots of speed-tests (across providers) on online forums. We gather these screenshots for further analysis while getting post data.
\end{itemize}
\vspace{-0.1in}

\vspace{-0.1in}
\section{Events drive sentiment}
\label{sec:event_sentiment}
\vspace{-0.05in}

Redditors' sentiments on \starlinksubreddit{} are influenced by events. We systematically identify peaks of strong sentiment and automate the discovery of related events of interest. 

\begin{figure*}[tbh]
  \centering
  \subfigure[]{\label{fig:sentiment_temporal}
    \includegraphics[height=4.8cm]{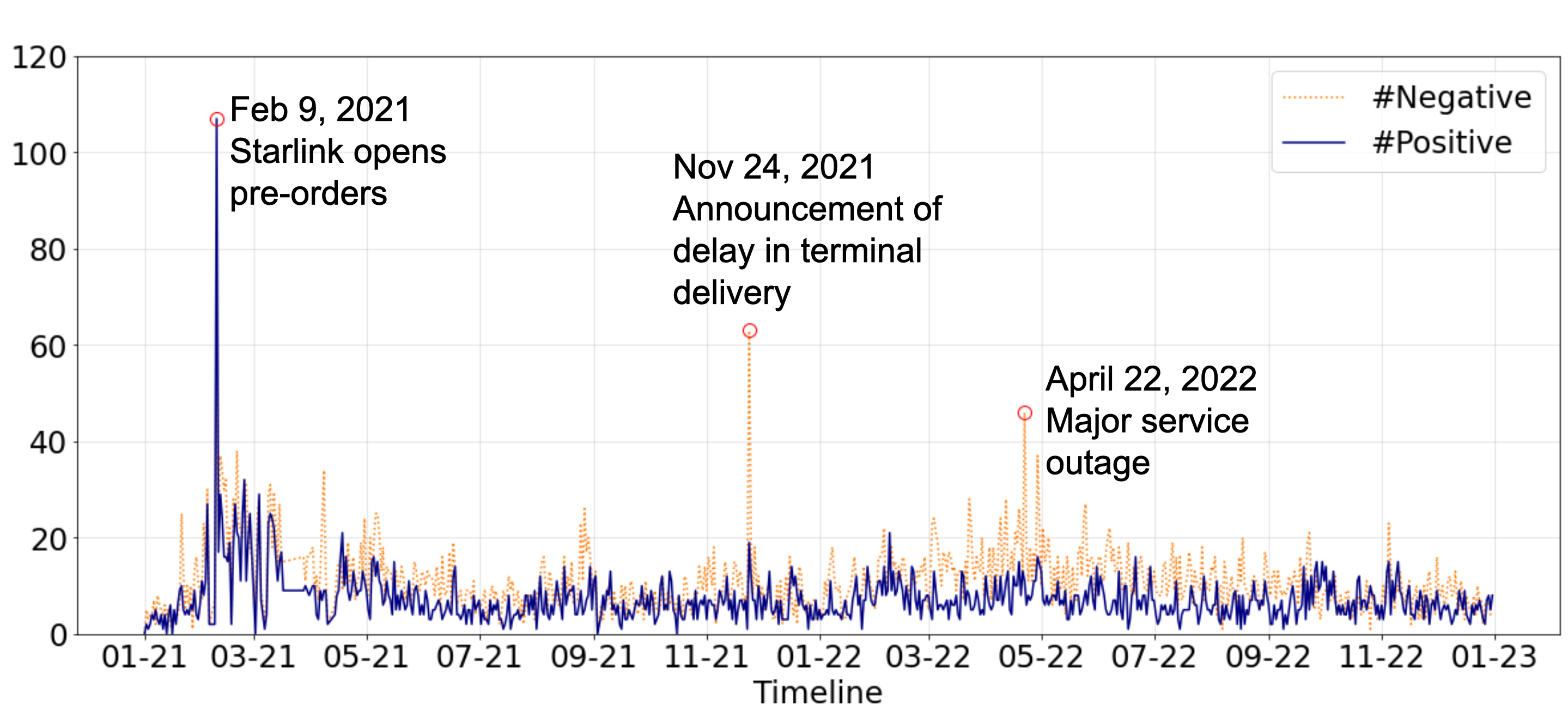}}
  \subfigure[]{\label{fig:outage_wordcloud}
    \includegraphics[height=4.5cm]{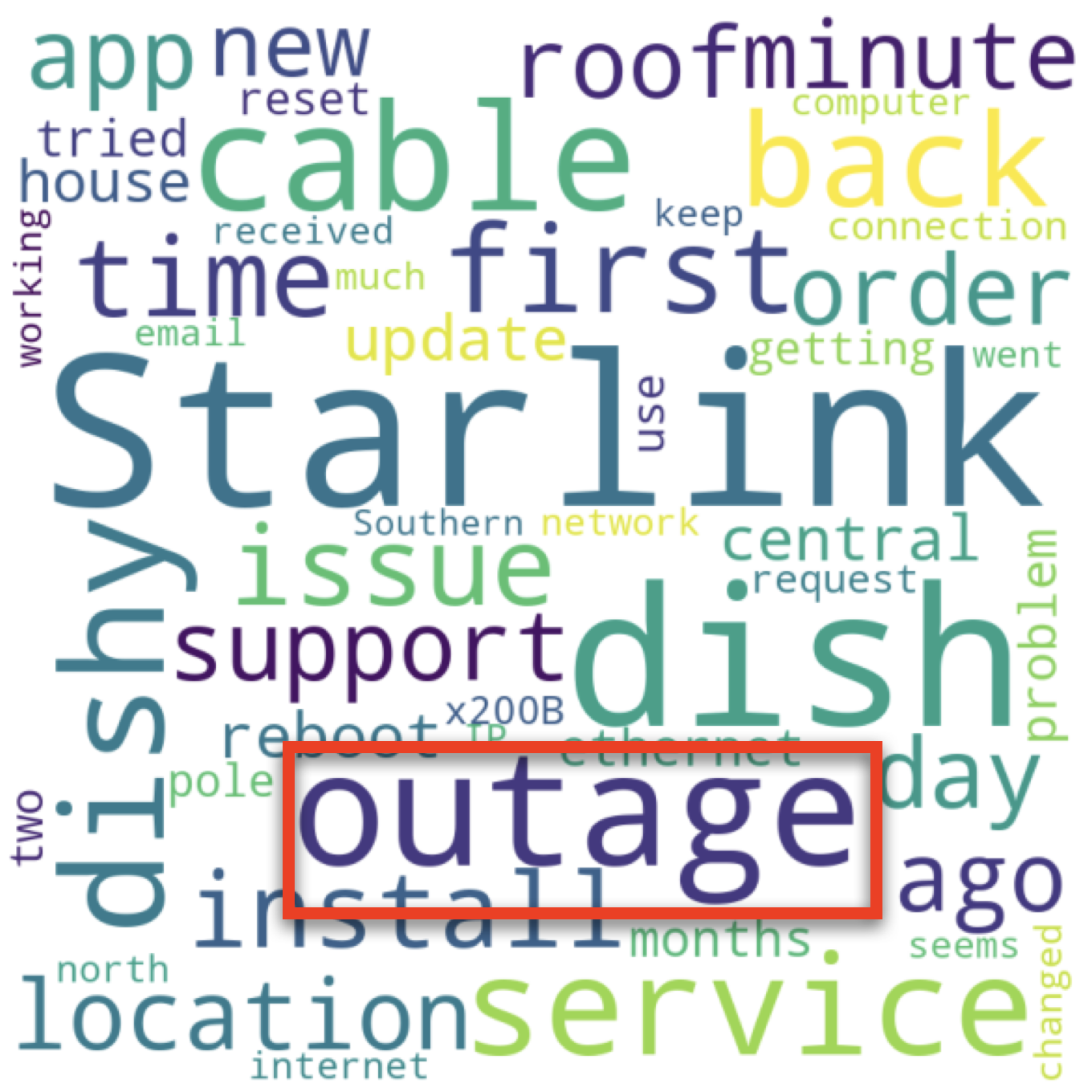}}
  \vspace{-0.15in}
  \caption{Although \starlinksubreddit{} is an active subreddit, (a) only a few sentiment peaks are observed. We pick the top $3$ peaks and tie them to actual events that led to these peaks. 
  (b) shows the word cloud for the $3^{rd}$ highest peak observed during and after a large-scale service outage.}
  \label{fig:sentiment}
\end{figure*}

For each day between Jan'$21$ and Dec'$22$, we analyze the sentiment of individual post content (text) using Azure's Cognitive Services.
The sentiment analysis service assigns $3$ different scores -- positive, negative, and neutral -- to each piece of text, which add up to $1$. We count the number of posts with strong positive ($\geq${}$0.7$) and negative ($\geq${}$0.7$) scores per day and plot them in Fig.~\ref{fig:sentiment_temporal}.

Our framework could discover sentiment peaks (day-wise bins), generate word clouds (using NLTK) across all posts and comments over a day for those peaks, and discover relevant news articles by searching online for the keywords (top $3$ uni-grams) appended with `Starlink' for the custom date. This pipeline enables the framework to annotate sentiment peaks with events that drive them. 

The top $3$ sentiment peaks, as shown in Fig.~\ref{fig:sentiment_temporal}, correspond to events of $3$ distinct flavors. On $9^{th}$ Feb'$21$, Redditors showed strong positive sentiment towards Starlink opening up pre-ordering of user terminals in the US, Canada, and UK~\cite{starlink_preorder}. On $24^{th}$ Nov'$21$, SpaceX's email~\cite{starlink_delay} to pre-order customers on delay in terminal delivery led to a negative sentiment peak. Fig.~\ref{fig:outage_wordcloud} shows the word cloud corresponding to the third highest peak ($22^{nd}$ Apr'$22$) which is driven by negative sentiment. The $3^{rd}$ most common word in the generated word cloud is `outage'. Interestingly, we could not find any relevant news though for this date, although Redditors from $14$ different countries (including $\sim${}$190$ reports from the US) confirmed an outage online. 

On further exploration, we could detect more such sentiment peaks which correspond to similar service outages, many of which were reported by the media~\cite{starlink_outage1,starlink_outage2,starlink_outage3}. Across such outages, negative sentiment (number of strongly negative posts and comments on outage as a fraction of all related posts and comments) varies by as much as $70\%$ depending on the nature of the outage. While some outages lasted for tens of minutes to hours, the one event that generated the most negative sentiment was on $10^{th}$ Mar'$21$ -- during the early beta-testing phase. Redditors reported~\cite{reddit_outage_thread_10Mar21} multiple short outages resulting in frequent call drops throughout the day, thus adding to their annoyance.

\begin{figure}[tb]
        \begin{center} \includegraphics[width=\columnwidth]{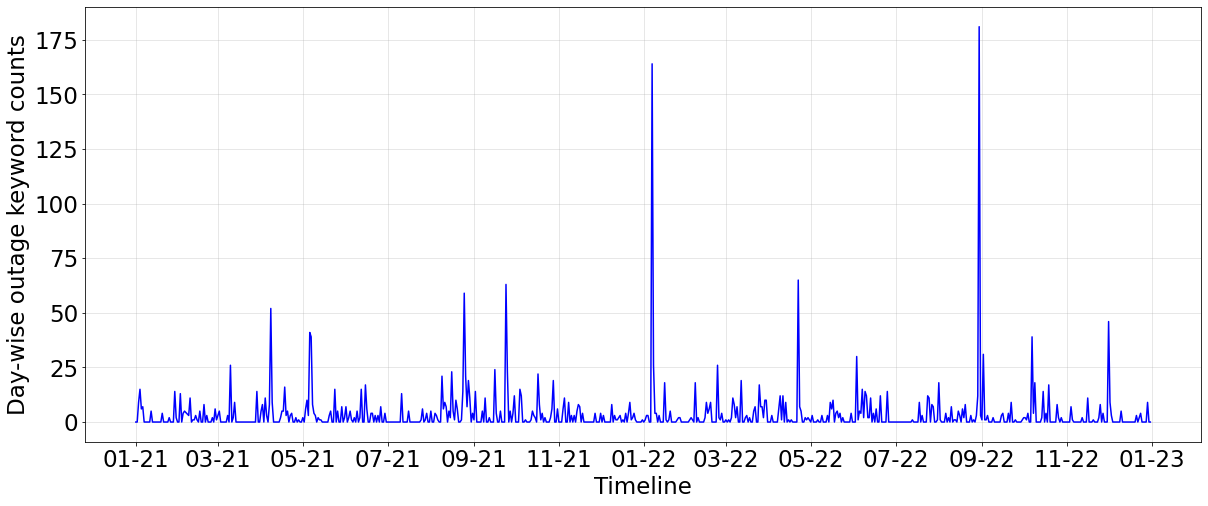}
        \vspace{-0.1in}
                \caption{While a few larger outages sparked a lot of discussions on \starlinksubreddit{}, outages with smaller impacts are quite frequent. Threads with positive or neutral sentiments have been filtered out.
                }
                \label{fig:outage_temporal}
        \end{center}
        \vspace{-0.2in}
\end{figure}

\begin{figure*}[t]
  \centering
  \subfigure[]{\label{fig:post_popularity}
\includegraphics[width=0.3\textwidth]{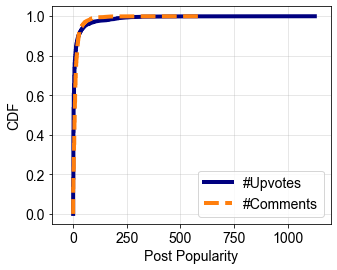}}
  \subfigure[]{\label{fig:roaming_wordcloud}    \includegraphics[width=0.65\textwidth]{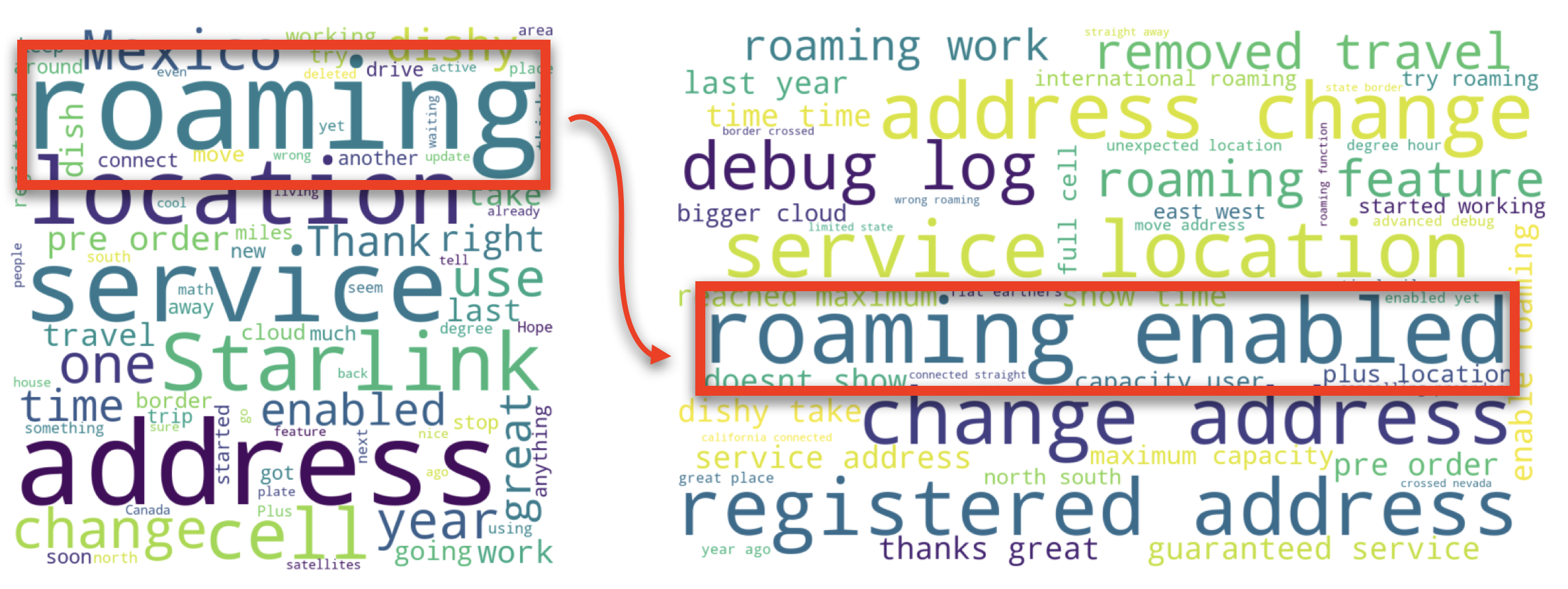}}
  \vspace{-0.15in}
  \caption{(a) Popularity of posts has a long tail with a few posts attracting many comments and up-votes. For Feb'$21$, the second most popular post (published on $23^{rd}$ Feb) has (b) \textbf{roaming} and \textbf{roaming enabled} as the most common uni-gram and bi-gram respectively.
  }
  \label{fig:roaming}
\end{figure*}

While the above approach (step $1$) is aimed at identifying sentiment peaks tied to various events, both networking and otherwise, we dive deeper and focus more on network service outages next (step $2$). We created a library of uni/bi-grams which are common in `outage' posts and comments identified following the above procedure. We manually verified that keywords in this library are meaningful in the specific context and got rid of outliers. We then identified threads with such keywords and negative sentiments (i.e., we filtered out threads with such keywords but positive or neutral sentiments) assuming outages to be `negative' events. Fig.~\ref{fig:outage_temporal} plots the day-wise frequencies of these keywords in such \textsl{negative} sentiment threads. $7^{th}$ Jan'$22$ and $30^{th}$ Aug'$22$ have the largest spikes of such keywords and correspond to large reported outages~\cite{starlink_outage7Jan22,starlink_outage30Aug22}. But more interestingly, there are numerous shorter peaks over time which correspond to local transient outages. Interestingly, in Fig.~\ref{fig:sentiment_temporal} we did not observe any sentiment peaks for the above $2$ dates when there were global outages because Redditors were also discussing other non-networking positive events that coincided with these network outages. The $2$-step process (leveraging specific keywords library and sentiment analysis) helps us zoom in on the particular event of interest -- network outage in this case.

OOkla's Downdetector~\cite{downdetector} also uses sentiment data from Twitter alongside direct reports, albeit the detailed methodology is not publicly available. Our framework, on the other hand, leverages a 2-step process, as detailed above, that could be used to analyze both networking and non-networking events of interest. Social platform-based methods could detect transient outages (probably due to network updates, misconfigurations, etc.) which might go undetected/unreported otherwise. In these initial days of LEO broadband offerings, understanding such transient downtimes is crucial in fixing performance bottlenecks and misconfigurations.

IODA~\cite{ioda_outage} relies on active probing and BGP data to detect Internet outages. While the methodology works well for the terrestrial Internet, it would be interesting to revisit the same in the context of LEO networks -- LEO terminals in a region might transiently have no satellites within the field of view and experience temporary downtime. We keep a deeper comparison with IODA for future work.

\vspace{-0.1in}
\section{Know before we know!}
\label{sec:roaming}
\vspace{-0.05in}

What are some of the most popular topics being discussed on \starlinksubreddit{} and what could we infer from them? For each month, we identify the list of popular posts and explore the topics of discussion. To do this, we rely on $2$ metrics -- the number of upvotes and the number of comments per post. Fig.~\ref{fig:post_popularity} plots the corresponding CDFs for the month of Feb'$22$. We observe that both the CDFs have long tails -- we pick popular posts which lie at the intersection of the $99^{th}$ percentiles of both the CDFs. For Feb'$22$, there are $1$,$898$ posts in total, and $8$ of them are identified to be popular based on the definition above. 

The most popular post is on the Ukraine war and the role of SpaceX Starlink in providing connectivity. We rather focus on the second most popular post which has $500+$ upvotes and $100+$ comments. We use NLTK to remove common stop words 
and create a word cloud of uni-grams for the post and the  comments on the post. Fig.~\ref{fig:roaming_wordcloud} shows that the most common uni-gram being discussed is `roaming'. As a next step, we generate a similar word cloud but this time with the bi-grams. The most common bi-gram is `roaming enabled'. We also analyze the sentiment scores of the specific post and the comments, and found them to be primarily \textsl{positive}, which increases our confidence on the fact that Starlink roaming might have been enabled in early February, $2022$.

Interestingly, while Redditors discussed Starlink roaming being functional on $23^{rd}$ Feb'$22$ already (we could also identify a smaller peak on the same topic on $15^{th}$ Feb), SpaceX officially notified~\cite{starlink_portability} Starlink subscribers on portability (roaming) mode only in May, $2022$ -- $3$ months down the line. Elon Musk, CEO of SpaceX, tweeted earlier~\cite{musk_roaming} on $3^{rd}$ March on mobile roaming being enabled, still $\sim${}$2$ weeks after we could detect the earliest signal from \starlinksubreddit{}. So, our framework could automatically detect that roaming have been enabled on the Starlink network way before any public announcements from SpaceX on the same.

Our methodology above could be used for different use cases. For example, ISPs could use such a tool to keep track of competition and be on the forefront of innovation. Also, while exploring the subreddit manually, we observed posts~\cite{reddit_link1,reddit_link2,reddit_link3,reddit_link4} which discuss deep technical aspects.
An extension to our methodology could allow ISPs to pinpoint such public discussions and learn from community expertise. As social networks get all the more ingrained in our lives, and with more people becoming comfortable discussing ideas and reporting facts online, ISPs could probably benefit from tapping into the public whiteboard discussions on social platforms. 

Also, at these early stages of LEO broadband, automatic identification of such events will allow potential customers, researchers, and Internet enthusiasts to keep track of all the features and their evolution over time. A potential customer in a remote area who `needs' the roaming capabilities might benefit from such early information on the same.
\aryan{We also observed Redditors discussing~\cite{orion1,orion2,orion3,orion4,orion5,orion6,orion7} performance bottlenecks (and opportunities) of applications, like video conferencing, over LEO broadband. These social media insights could allow application providers to pinpoint issues over these new networks and fix them early toward offering improved experience.}

\vspace{-0.1in}
\section{Following the shifting fulcrum}
\label{sec:bandwidth}
\vspace{-0.05in}

Social media users share photos and screenshots while posting online. Redditors on \starlinksubreddit{} often share screenshots (or links to them) of network performance test reports to spark discussions. We gather all such test report screenshots across test providers like Ookla~\cite{ookla_speedtest}, Fast (powered by Netflix)~\cite{fast_speedtest}, Starlink itself, and others, and extract uplink speed, downlink speed, latency information, etc. using Azure's Optical Character Recognition (OCR)~\cite{azure_ocr}. After applying a set of heuristics, as discussed in \S\ref{sec:methods}, to rule out false positives, we could identify $\sim${}$1$,$750$ reports of Starlink speed-tests being shared publicly between Jan, $2021$ and Dec, $2022$. In this section, we focus on the evolution of `observed' downlink speed during this period and users' perceptions of the same.

\begin{figure*}[tbh]
        \begin{center} \includegraphics[width=\textwidth]{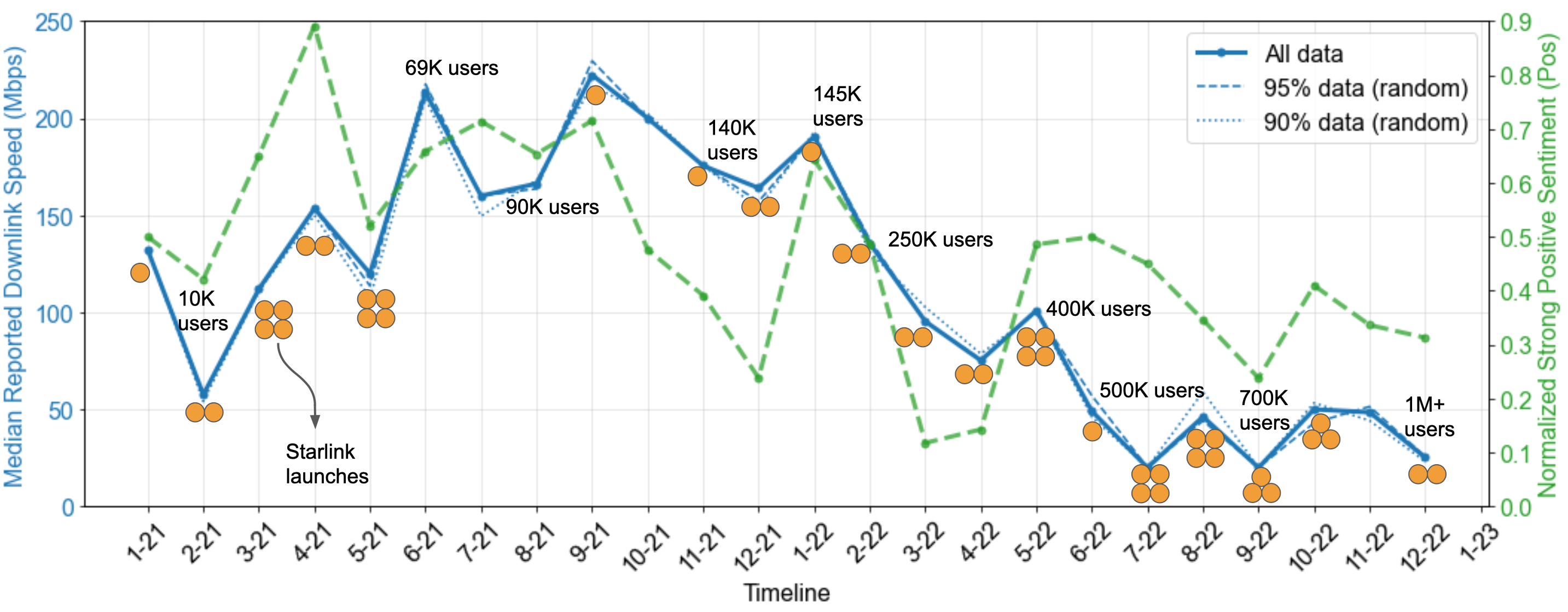}
                \vspace{-0.15in}
                \caption{Observed downlink speeds on Starlink evolve with more launches and customers. User sentiment largely follows observed speed. Number of Starlink users, as and when available as public information, has been used to annotate the plot.}
                \label{fig:downlink_evolve}
        \end{center}
\end{figure*}

\parab{Demand vs. supply}
Fig.~\ref{fig:downlink_evolve} shows the change in observed downlink speeds with time. For each month, we plot the median speeds across all shared screenshots of Starlink speed tests. We annotate the observed speeds with the number of Starlink launches~\cite{starlink_launches} and also the reported number of Starlink users (whenever public information is available)~\cite{beta_10K,beta_69K,beta_90K,beta_140K,beta_250K,beta_400K,beta_500K} during a month. We also plot the monthly median downlink speeds with $95\%$ and $90\%$ of the monthly speed data picked uniformly at random -- the plots closely follow each other showing that the observed medians are considerably stable.

We observe that, between Jan and Sep'$21$, the median downlink speeds increased in general. SpaceX launched ($\sim${}$60$ satellites per launch) $14$ times during this period and the number of users increased from $10K$ (in Feb) to $90K$ (in Aug).  Further, between Sep'$21$ and Dec'$22$, there has been an almost steady decrease in observed speeds although SpaceX launched batches of Starlink satellites $32$ times. Note, however, that the number of reported Starlink users increased from $90K$ to $1M$ (and beyond) during the same period, resulting in more than $1000\%$ rise in downlink demand, assuming a linear increase in demand with users. Note that between Nov'$21$ and Jan'$22$, there were $4$ Starlink launches while the number of users increased by only $5K$ (announcement by Starlink on delay in terminal delivery in Nov'$21$, as shown in Fig.~\ref{fig:sentiment_temporal}). This resulted in slight improvements in observed bandwidth in Jan'$22$.
In March, April, and June'$21$, observed downlink speeds improved as new satellites were deployed steadily. But it is not always a straightforward reflection of deployment as more Starlink users are also added continually, as is evident in some other cases. On the contrary, if there aren't any new launches, and new users are added, the observed speeds decrease. Between Jun and Aug'$21$, $21K$ new users started using Starlink with no new launches happening. This is reflected in the sharp decrease in median speeds during the period. Beyond Sep'$21$, reported bandwidths have decreased almost steadily given the large increase in demand as Starlink service expanded to various countries across the globe.

The broad observation here is that more Starlink launches do not always result in higher observed bandwidth, although the aggregate bandwidth of the Starlink system should be increasing with the addition of satellites. It is a complex calculus involving both supply and demand, and it is always a race to add more satellites and cater to the ever-increasing aggregate bandwidth demands.

We did not quantify any bias inherent in such social media-based estimates. Note that unbiased absolute values of download speeds are not critical for our analyses; we rather needed relative changes in download speeds and corresponding sentiments (discussed next) from one month to the next. As part of future work, any bias arising due to demographics in a social community will hopefully reduce as we span across more social platforms. We discuss this in further depth under \S\ref{sec:discussions}.

\parab{The wheel of time}
With time, the perception of users on network performance changes. In Fig.~\ref{fig:downlink_evolve} we also plot (green, dashed) the strong positive sentiment of users on downlink speeds. To do this, we analyze the sentiment of posts (text content) that share Starlink speed-test reports using Azure's Cognitive Services. We identify posts with strong positive ($\geq${}$0.7$) or negative ($\geq${}$0.7$) scores, and define the normalized strong positive sentiment score ($Pos$) as the number of strong positive posts over sum of the number of strong positive and negative posts in a month thus filtering out edge cases when identifying the sentiment is hard.

We observe that $Pos$ broadly follows the observed downlink speed trends, but there are interesting exceptions. For instance, while downlink speed is higher in Dec'$21$ than Apr'$21$, $Pos$ is drastically lower for Dec'$21$. We believe this is because user sentiment is, in general, a reflection of both short-term and long-term conditioning -- users get acclimatized to their current network conditions and give negative sentiment for any degradation in network conditions even if such conditions are better than the past. The exact inverse of this trend is also visible -- the downlink speeds decrease between Mar'$22$ and Dec'$22$ while the $Pos$ improves over time. This demonstrates users getting conditioned to lower speeds, but not necessarily attachment and loyalty to the ISP. While we observed users frequently discussing application performance over LEO networks on social platforms, we keep such an in-depth and finer-granularity study of observed bandwidth versus user sentiment for future work.

\section{Broader implications}
\label{sec:applications}

While we have focused on SpaceX Starlink and the corresponding subreddit \starlinksubreddit{} in this preliminary study, our methodology and framework could span multiple online social communities and LEO providers to collect interesting insights in the absence of large-scale measurement platforms. Even beyond LEO, such a framework allows us to revisit our perception of Internet measurement.

\parab{User-centric} For relatively new offerings like LEO broadband Internet, it is critical to assess user perception and engagement to gauge the market demand and dynamics. How do the customers react and express themselves online in response to broad events such as infrastructure upgradation, outage, peering and partnership announcements, etc.? An NLP and vision services-enabled framework like this could potentially tap into the large corpus of user feedback, discussions, and debate publicly available on various online social platforms. 

\parab{Complementarity to measurement testbeds} While platforms like PlanetLab~\cite{planetlab} and M-Lab~\cite{MLab} capture Internet measurements at scale over years, LEO broadband networks with all the uniqueness and orbital dynamics also need to be measured at such global scales. But even if such an LEO measurement platform comes into existence, insights from social media capturing user perception and sentiment could complement such measurements, offering a more holistic view of these networks at the early stages of deployment.

\parab{Active user engagement} One could take a step further and engage actively with users on social platforms. It should be straightforward to share back the insights at both aggregate and granular levels with users. For example, when a user uploads a speed test screenshot, the framework could employ bots that allow the user to compare their experience with that of others. A user can opt out of such a service at will and could also request the framework to delete their data. If a user rather wants to opt in, there could be different levels of user engagement -- allowing the framework to collect more data (e.g. physical coordinates of user terminals) and run more speed tests.
Enthusiastic users could also be willing to share some of their resources for running arbitrary network experiments on a slice. A global LEO measurement testbed would need spatiotemporal diversity, given the unique dynamicity and geometry of the constellations, and benefit from such large-scale voluntary participation.

\section{Discussions}
\label{sec:discussions}

\parab{AI for user sentiment} Generative AI models~\cite{gpt3, llama,openai2023gpt4} could effectively summarize and quantify user feedback and sentiment while removing harmful/biased content. Custom AI models that could understand the user pulse on social media at scale could augment the insights generated using cutting-edge vision and sentiment analysis tools.

\parab{The social network behavioral bias:} We did not analyze the bias that might occur due to a divergence between the experience of users and what they choose to report on social networks. For example, a user might be running network speed tests multiple times and reporting only the better/worse ones. Another bias might arise due to users in a public group being over-enthusiastic about the topic. For example, in the case of Reddit, discussions on \starlinksubreddit{} could have a stronger negative sentiment toward Starlink's competitors. Also, bias might arise due to the socio-demographics of users active on a particular social platform~\cite{hargittai2020potential}. As part of future work, we plan to expand across multiple social platforms and communities and combine social media data with other sources of information to quantify/mitigate this bias, if existent.

\parab{Gaming the framework:} Different stakeholders including users and ISPs could game the framework. For example, automated bots on Reddit could publish posts with strong sentiments to skew the aggregate analyses. We could rely on ongoing research on detecting bot-generated text and bot behavior~\cite{alothali2018detecting,schuchard2019bot,shi2019detecting,beskow2018bot} to tackle this problem. AI/ML-based techniques~\cite{luceri2020detecting,al2022online} to detect large-scale trolls on social media could also be plugged in to make the framework more robust to such gaming. One problem though is that on a specific social community, like \starlinksubreddit{}, the number of interactions is not too high volume, thus making it vulnerable to gaming. Nevertheless, gathering data across multiple platforms and communities might address this issue.

\parab{Tackling ethical and privacy concerns:} While we do not use user identification information in this work, one might extend the system to engage more closely with users, as discussed in \S\ref{sec:applications}. In such cases, we should carefully consider the ethical and privacy concerns that might arise while dealing with user-level data. Data should be anonymized, and user consent should precede active user engagement. One should also have a mechanism in place to permanently delete user data upon request, conforming to privacy regulations such as GDPR~\cite{gdpr}, CCPA~\cite{ccpa}, etc.

\parab{The more the better:} In this preliminary study, we only analyze data from a single platform (Reddit) and for a single ISP (Starlink). The framework could be extended to also analyze other social platforms such as Facebook, Twitter, Discord, etc., and other ISPs. The more data and the more diverse group culture explored, the easier it would be to identify sentiment peaks and important network (and non-network) events. Also, more data across groups with related but slightly different agendas might reduce the behavioral bias discussed above.

\parab{Location-based analysis (LEO specific)} More data on speed-test reports across test providers and social platforms could allow us to analyze the impact of constellations' geometry, Earth's shape, cell congestion, interference with GEO satellites, etc. on Starlink performance. This might require users to engage more with the framework and share explicit location information.

\section{Related work}

Recent breakthrough advances in natural language processing~\cite{devlin2018bert,dai2019transformer,yang2019xlnet,liu2019roberta,qi2020stanza,azure_acs,google_NL_AI,openai2023gpt4} and computer vision~\cite{girshick2014rich,redmon2016you,he2017mask,brock2018large,karras2019style} have made text and image scanning, parsing, and comprehension capabilities~\cite{luan2019general,huang2019icdar2019,liu2019graph} ubiquitous among other things. Sentiment analysis, which is used for detecting sentiments underlying text, has been exhaustively applied to a multitude of use cases -- understanding product feedback and preferences~\cite{liu2012sentiment,anto2016product}, and predicting trends and outcomes of large-scale events~\cite{matalon2021using,karalevicius2018using,bermingham2011using} leveraging the corpus of publicly available user interactions on online social platforms. Leveraging social media to understand networks better is not a completely new space -- sentiment data have been used to understand/detect mobile network performance~\cite{qiu2010listen,hsu2011using} and demands~\cite{yang2016estimating}, service availability and failures~\cite{motoyama2010measuring,takeshita2015early}, and attacks and security vulnerabilities~\cite{al2013leveraging,ritter2015weakly,sabottke2015vulnerability,khandpur2017crowdsourcing,chambers2018detecting,shu2018understanding}. Motivated by this large body of past work, we explore a similar methodology to understand SpaceX Starlink service leveraging crowdsourced information available on social media.

While many measurement platforms~\cite{planetlab,MLab,ripe_atlas,ookla_speedtest,fast_speedtest} exist that gather Internet performance metrics at scale, we did not find a measurement platform yet targeted to analyze the nuances of LEO dynamics -- temporal changes in latency and bandwidth independent of the congestion, impact of the constellation geometry, weather, and interference with other similar services, etc. While the research community has already engaged in simulations~\cite{kassing2020exploring,bhattacherjee2019network,handley2018delay} and limited measurements~\cite{starlink_perf_firstlook,leo_geo_aus} of these networks, measurements at scale are still unavailable. A recent work~\cite{starlink_browserside} on gathering LEO measurements leverages a custom-built browser plugin -- our work is complementary to this work in the absence of large-scale LEO measurement platforms. Even if large-scale LEO measurement platforms become available in the future, our framework has the potential to complement such platforms by offering a user-centric view of the services. A more recent work~\cite{pan2023measuring} uses Reddit to ``crosscheck'' Starlink measurement insights, but we rather consume public data on the \starlinksubreddit{} subreddit as the primary source of performance and user sentiment measurements.

\section{Conclusion}

We leverage recent advances in language and vision capabilities to mine social media and understand users' perceptions of SpaceX Starlink network performance and events. This framework could complement LEO broadband measurement tools by offering a user-centric view of these networks. Our framework could detect important network events days/weeks before public announcements, detect sentiment peaks and related events (like outages), and understand the evolving perception of bandwidth on the SpaceX Starlink network.

%
%
%
%
\bibliographystyle{splncs04}
\bibliography{pam22} 
\end{document}